\pdfoutput=1

\documentclass[journal]{IEEEtran}
\usepackage[utf8]{inputenc}

\usepackage{cite}
\usepackage[english]{babel}
\ifCLASSINFOpdf
	\usepackage[pdftex]{graphicx}
	\usepackage{graphics}
\else
  	\usepackage[dvips]{graphicx}
  	\DeclareGraphicsExtensions{.eps}
\fi
\usepackage{rotating}

\usepackage{amsmath}
\usepackage{amssymb}
\usepackage{array}
\usepackage{mdwmath}
\usepackage{mdwtab}
\usepackage[linesnumbered,ruled,vlined]{algorithm2e}
\DontPrintSemicolon
\usepackage{algorithmic}
\SetCommentSty{mycommfont}

\SetKwInput{KwInput}{Input}                % Set the Input
\SetKwInput{KwOutput}{Output}              % set the Output

\usepackage{array}
\ifCLASSOPTIONcompsoc
  \usepackage[caption=false,font=normalsize,labelfont=sf,textfont=sf]{subfig}
\else
  \usepackage[caption=false,font=footnotesize]{subfig}
\fi

\usepackage{stfloats}
\usepackage{url}
\usepackage{color,soul}
\usepackage{nomencl}

\makenomenclature
\usepackage{etoolbox}
\renewcommand\nomgroup[1]{%
	\ifstrequal{#1}{I}{\vspace{10pt} \item[\textbf{Indices and Sets}]}{%
	\ifstrequal{#1}{P}{\vspace{10pt} \item[\textbf{Parameters}]}{%
	\ifstrequal{#1}{Q}{\vspace{10pt} \item[\textbf{Parameters ADMM}]}{
	\ifstrequal{#1}{V}{\vspace{10pt} \item[\textbf{Variables}]}{
	\ifstrequal{#1}{A}{\item[\textbf{Abbreviations}]}}{}}
}%
}
}
\usepackage{eurosym}

\usepackage{enumitem}
\setlist{leftmargin=4.5mm}

\begin{document}
%
% paper title
% Titles are generally capitalized except for words such as a, an, and, as,
% at, but, by, for, in, nor, of, on, or, the, to and up, which are usually
% not capitalized unless they are the first or last word of the title.
% Linebreaks \\ can be used within to get better formatting as desired.
% Do not put math or special symbols in the title.
%\title{Two-steps Fast-PJ-ADMM Distributed Flexibility Services Provision for Local Markets}
%\title{Centralised and Distributed Aggregated Flexibility Services Provision Algorithms for Local Markets}
\title{Centralised and Distributed Optimization for Aggregated Flexibility Services Provision}
%
%
% author names and IEEE memberships
% note positions of commas and nonbreaking spaces ( ~ ) LaTeX will not break
% a structure at a ~ so this keeps an author's name from being broken across
% two lines.
% use \thanks{} to gain access to the first footnote area
% a separate \thanks must be used for each paragraph as LaTeX2e's \thanks
% was not built to handle multiple paragraphs
%

\author{Pol~Olivella-Rosell,~\IEEEmembership{Student Member,~IEEE,}
        Francesc~Rul$\cdot$lan,
        Pau~Lloret-Gallego,
        Eduardo~Prieto-Araujo,~\IEEEmembership{Member,~IEEE,}
        Ricard~Ferrer-San-Jos\'{e},~\IEEEmembership{Student Member,~IEEE,}
        Sara~Barja-Martinez,
        Sigurd~Bjarghov,~\IEEEmembership{Student Member,~IEEE,}
	    Venkatachalam~Lakshmanan,
       	Ari~Hentunen,
       	Juha~Forsstr\"{o}m,
        Stig~\O{}degaard~Ottesen,
		Roberto~Villafafila-Robles,~\IEEEmembership{Member,~IEEE,}
		and Andreas~Sumper,~\IEEEmembership{Senior Member,~IEEE}% <-this % stops a space
%\thanks{P.Olivella-Rosell, P.Lloret-Gallego, E.Prieto-Araujo, R.Ferrer-San-Jose, S.Barja-Martinez, R.Villafafila-Robles and A.Sumper are with the CITCEA-Department
%of Electrical Engineering, Universitat Politecnica de Catalunya, Barcelona,
%Spain e-mail: pol.olivella@outlook.com}% <-this % stops a space
%\thanks{F.Rul$\cdot$lan is with University College London}% <-this % stops a space
%\thanks{S.\O.Ottesen is with eSmart Systems AS}
%\thanks{S.Bjarghov and V.Lakshmanan are with NTNU}
%\thanks{A.Hentunen and J.Forsstr\"{o}m are with VTT}
\thanks{P. Olivella-Rosell, P. Lloret-Gallego, E. Prieto-Araujo, R. Ferrer-San-Jose, S. Barja-Martinez, R. Villafafila-Robles and A. Sumper are with  CITCEA, Universitat Politecnica de Catalunya, Barcelona,
	Spain e-mail: (pol.olivella@outlook.com). F. Rul$\cdot$lan is with University College London, S. \O. Ottesen is with eSmart Systems AS, S. Bjarghov and V. Lakshmanan are with NTNU, A. Hentunen and J. Forsstr\"{o}m are with VTT.}}
%\thanks{Manuscript received July 1, 2019?; revised...}}

% note the % following the last \IEEEmembership and also \thanks - 
% these prevent an unwanted space from occurring between the last author name
% and the end of the author line. i.e., if you had this:
% 
% \author{....lastname \thanks{...} \thanks{...} }
%                     ^------------^------------^----Do not want these spaces!
%
% a space would be appended to the last name and could cause every name on that
% line to be shifted left slightly. This is one of those "LaTeX things". For
% instance, "\textbf{A} \textbf{B}" will typeset as "A B" not "AB". To get
% "AB" then you have to do: "\textbf{A}\textbf{B}"
% \thanks is no different in this regard, so shield the last } of each \thanks
% that ends a line with a % and do not let a space in before the next \thanks.
% Spaces after \IEEEmembership other than the last one are OK (and needed) as
% you are supposed to have spaces between the names. For what it is worth,
% this is a minor point as most people would not even notice if the said evil
% space somehow managed to creep in.

% The paper headers
\markboth{Submitted to Journal of Transactions on Smart Grid}%
{Shell \MakeLowercase{\textit{et al.}}: Bare Demo of IEEEtran.cls for IEEE Journals}
% The only time the second header will appear is for the odd numbered pages
% after the title page when using the twoside option.
% 
% *** Note that you probably will NOT want to include the author's ***
% *** name in the headers of peer review papers.                   ***
% You can use \ifCLASSOPTIONpeerreview for conditional compilation here if
% you desire.

% If you want to put a publisher's ID mark on the page you can do it like
% this:
%\IEEEpubid{0000--0000/00\$00.00~\copyright~2015 IEEE}
% Remember, if you use this you must call \IEEEpubidadjcol in the second
% column for its text to clear the IEEEpubid mark.

% use for special paper notices
%\IEEEspecialpapernotice{(Invited Paper)}

% make the title area
\maketitle

% As a general rule, do not put math, special symbols or citations
% in the abstract or keywords.
\begin{abstract}
The recent deployment of distributed battery units in prosumer premises offer new opportunities for providing aggregated flexibility services to both distribution system operators and balance responsible parties.
The optimization problem presented in this paper is formulated with an objective of cost minimization which includes energy and battery degradation cost to provide flexibility services.
A decomposed solution approach with the alternating direction method of multipliers (ADMM) is used instead of commonly adopted centralised optimization to reduce the computational burden and time, and then reduce scalability limitations. 
In this work we apply a modified version of ADMM that includes two new features with respect to the original algorithm: 
first, the primal variables are updated concurrently, which reduces significantly the computational cost when we have a large number of involved prosumers; 
second, it includes a regularization term named Proximal Jacobian (PJ) that ensures the stability of the solution.
A case study is presented for optimal battery operation of 100 prosumer sites with real-life data. 
The proposed method finds a solution which is equivalent to the centralised optimization problem and is computed between 5 and 12 times faster. 
Thus, aggregators or large-scale energy communities can use this scalable algorithm to provide flexibility services.
%The proposed method finds a solution which is equivalent to the centralised optimization problem and it goes between 5 and 12 times faster in the case study. 

\end{abstract}

% Note that keywords are not normally used for peerreview papers.

% For peer review papers, you can put extra information on the cover
% page as needed:
% \ifCLASSOPTIONpeerreview
% \begin{center} \bfseries EDICS Category: 3-BBND \end{center}
% \fi
%
% For peerreview papers, this IEEEtran command inserts a page break and
% creates the second title. It will be ignored for other modes.
\IEEEpeerreviewmaketitle

%\section*{Nomenclature}
%\addcontentsline{toc}{section}{Nomenclature}

\nomenclature[I]{$I$}{Set of prosumer sites, indexed by $i$}
\nomenclature[I]{$T$}{Set of periods/time slots in the planning horizon, indexed by $t$}
\nomenclature[I]{$T^\pm$}{Subset of periods with up (+) or down (-) regulation}
%Site parameters
\nomenclature[P]{$P_{i,t}^{buy}$}{Price at energy part for buying electricity in period $t$ in site $i$ [EUR/kWh]}
\nomenclature[P]{$P_{i,t}^{sell}$}{Price at energy part for buying electricity in period $t$ in site $i$ [EUR/kWh]}
\nomenclature[P]{$P_{i,t}^{gen}$}{Penalty cost for curtailing photovoltaic energy production [EUR/kWh] of site $i$ in period $t$}
\nomenclature[P]{$X_{i}^{imp}$}{Maximum electricity import capacity of site $i$ per period [kWh]}
\nomenclature[P]{$X_{i}^{exp}$}{Maximum electricity export capacity of site $i$ per period [kWh]}
%Site variables
\nomenclature[V]{$\chi_{i,t}^{buy}$}{Amount of electricity bought in period $t$ by site $i$ [kWh]}
\nomenclature[V]{$\chi_{i,t}^{sell}$}{Amount of electricity sold in period $t$ by site $i$ [kWh]}
\nomenclature[V]{$\delta_{i,t}^{buy}$}{Binary variable=1 if site $i$ is importing electricity in period $t$, else 0}
\nomenclature[V]{$\delta_{i,t}^{sell}$}{Binary variable=1 if site $i$ is exporting electricity in period $t$, else 0}
%Aggregated parameters
\nomenclature[P]{$W_{t}^{flex} $}{Aggregated electricity consumption in period $t$ after meeting the flexibility request [kWh]}
\nomenclature[P]{$W_{t}^{base}$}{Aggregated baseline consumption in period $t$ [kWh]}
\nomenclature[P]{$FR_{t}$}{Flexibility request from DSO/BRP in period $t$ [kWh]}
\nomenclature[P]{$W_{i,t}^{l}$}{Inflexible load consumption in period $t$ in site $i$ [kWh]}
%Flexibility variables
\nomenclature[V]{$\psi_{i,t}$}{Amount of electricity produced from generating unit in period $t$ in site $i$ [kWh]}
\nomenclature[V]{$\sigma_{i,t}^{ch}$}{Amount of electricity charged to the battery unit in period $t$ in site $i$ [kWh]}
\nomenclature[V]{$\sigma_{i,t}^{dis}$}{Amount of electricity discharged from the battery unit in period $t$ in site $i$ [kWh]}
\nomenclature[V]{$\sigma_{i,t}^{soc}$}{Battery SOC of site $i$ in time step $t$ at the end of period $t$ [kWh]}
\nomenclature[V]{$\zeta_{i,t}^{flex}$}{Total cost for activating flexibility in period $t$ in site $i$ [EUR]}
\nomenclature[A]{FR}{Flexibility request}
\nomenclature[A]{PV}{Photovoltaic}
\nomenclature[A]{BRP}{Balance responsible party}
\nomenclature[A]{DSO}{Distribution system operator}
\nomenclature[A]{FD}{Flexibility device}
\nomenclature[A]{PTU}{Programing time unit}
\nomenclature[A]{SOC}{State-of-charge}
\nomenclature[A]{ADMM}{Alternating direction method of multipliers}
\nomenclature[A]{ALFO}{Aggregated level flexibility offer}
\nomenclature[A]{ALFM}{Aggregated level flexibility management}
\nomenclature[A]{LFM}{Local flexibibility market}
\nomenclature[A]{HEMS}{Home energy management system}
\nomenclature[A]{P2P}{peer-to-peer}
\nomenclature[A]{PJ}{Proximal Jacobian}
\nomenclature[A]{EU}{European Union}

\nomenclature[I]{$J$}{Set of segments for battery SOC, indexed by $j$}
\nomenclature[V]{$\sigma _{i,t}^{soc}$}{Battery SOC in time step $t$ in site $i$ [kWh]}
\nomenclature[V]{$\sigma _{i,t}^{ch}$}{Battery charging energy in time step $t$ in site $i$ [kWh]}
\nomenclature[V]{$\sigma _{i,t}^{dis}$}{Battery discharging energy in time step $t$ in site $i$ [kWh]}
\nomenclature[V]{$\sigma _{i,t,j}^{seg,soc}$}{Amount of electricity stored in the battery segment $j$ in time step $t$ in site $i$ [kWh]}
\nomenclature[V]{$\sigma _{i,t,j}^{seg,ch}$}{Battery charging energy in time step $t$ in segment $j$ in site $i$ [kWh]}
\nomenclature[V]{$\sigma _{i,t,j}^{seg,dis}$}{Battery discharging energy in time step $t$ in segment $j$ in site $i$ [kWh]}
\nomenclature[V]{$a^{inv,ch}(\cdot)$}{Battery inverter charging efficiency function [p.u.]}
\nomenclature[V]{$a^{inv,dis}(\cdot)$}{Battery inverter discharging efficiency function [p.u.]}
\nomenclature[V]{$\delta_{i,t}^{bat}$}{Binary variable=1 if battery of site $i$ is charged in period $t$, else 0}
\nomenclature[P]{$Q_i^{ch}$}{Maximum battery charging energy per time unit in site $i$ [kWh]}
\nomenclature[P]{$Q_i^{dis}$}{Maximum battery discharging energy per time unit in site $i$ [kWh]}
\nomenclature[P]{$A_i^{bat,ch}$}{Battery charging efficiency in site $i$ [p.u.]}
\nomenclature[P]{$A_i^{bat,dis}$}{Battery discharging efficiency in site $i$ [p.u.]}
\nomenclature[P]{$O_i^{min}$}{Minimum battery SOC in site $i$ [kWh]}
\nomenclature[P]{$O_i^{max}$}{Maximum battery SOC in site $i$ [kWh]}
\nomenclature[P]{$O_{i,j}^{max}$}{Maximum battery SOC in segment $j$ in site $i$ [kWh]}
\nomenclature[P]{$C_{i,j}$}{Marginal battery cycling ageing cost per segment $j$ in site $i$ [EUR/kWh]}
\nomenclature[P]{$S_i^{LT}$}{Amount of time intervals in battery lifetime in site $i$ [years]}
\nomenclature[P]{$W^{bat}_i$}{Battery voltage charging tuning factor in site $i$}

\nomenclature[Q]{$\tau^{incr}$}{Incremental parameter to accelerate penalty parameter $\rho$}
\nomenclature[Q]{$\tau^{decr}$}{Decremental parameter to decelerate penalty parameter $\rho$}
\nomenclature[Q]{$\lambda_{t}^{(k)}$}{Dual variable for constrained period $t$ in iteration $k$}
\nomenclature[Q]{$\rho^{(k)}$}{Penalty parameter in iteration $k$}
\nomenclature[Q]{$r_t^{(k)}$}{Primal error for constrained period $t$ in iteration $k$ [kWh]}
\nomenclature[Q]{$s_t^{(k)}$}{Dual error for constrained period $t$ in iteration $k$ [kWh]}
\nomenclature[Q]{$\gamma$}{Damping parameter}
\nomenclature[Q]{$K^{i}$}{Penalty for accumulated primal error}
\nomenclature[Q]{$K^{d}$}{Penalty for the dual error}
\nomenclature[Q]{$\epsilon^{pri}$}{Primal error threshold}
\nomenclature[Q]{$\epsilon^{dual}$}{Dual error threshold}
\nomenclature[Q]{$CT^{max}$}{Maximum computation time threshold}

%\printnomenclature[1.3cm]

%\begin{IEEEdescription}[\IEEEsetlabelwidth{$V_1,V_2,$}]
%	\item[\smash{\begin{IEEEeqnarraybox*}[][t]{l}
%			V_1,V_2,\\
%			\hphantom{V_1,{}}V_3
%	\end{IEEEeqnarraybox*}}] Three-phase PWM output line voltages.\\
%	\mbox{}
%	\item[\textbf{label}]
%	\item[$\theta$] Rotor angle (in ``electrical degrees'').
%	\item[$\omega$] Rotor (electrical) speed, corresponding to the time
%	derivative of $\theta$.
%\end{IEEEdescription}

\vspace{-0.3cm}
\section{Introduction}
\IEEEPARstart{I}{n} the context of smart grids and flexibility services in place, balance responsible parties (BRPs) and distribution system operators (DSOs) can benefit by activating flexibility in distribution grids. 
Electricity price volatility in Europe is increasing in electricity markets during scarcity periods, for example due to fluctuating generation. 
Additionally, distributed generation could compromise the stable operation of distribution grids and cause congestions.

In this scenario, remotely controlled distributed batteries at prosumer premises could help to provide flexibility services for dealing with the issues mentioned above, like INVADE H2020 project proposes \cite{INVADEWS}.
An energy cloud platform can remotely manage batteries and it can be operated by an aggregator with a portfolio formed by a group of prosumer sites.
These sites can belong to different BRPs and DSO, and they can be grouped according to their grid and BRP zones.
For instance, all sites with contracts of flexibility provision for DSO service within the same grid zone can be operated to respond to a DSO flexibility request (FR).
Therefore, DSOs can increase their grid capacity to host more renewable generation or reduce network congestions during peak production or consumption periods respectively.
Similarly, during the periods of high prices, BRPs could maintain their day-ahead generation and consumption portfolio by activating flexibility instead of paying penalties for their deviations or paying high intraday market costs to keep their energy portfolio \cite{VandenBerge2016}.
The present paper deals with the short-term operation of distributed batteries behind-the-meter in order to provide flexibility services to BRPs or DSOs at the minimum cost. 

\vspace{-0.35cm}
\subsection{Home energy management systems}
Previous work in topics of decision-making at site level and home energy management systems (HEMS) include optimization algorithms to find the best possible scheduling of flexibility devices (FDs). 
Therefore, HEMS algorithms consider that sites are independent of each other, metered separately and focuses on individual site's benefit. 
A detailed analysis of HEMS and FD is presented in \cite{Beaudin2015,Zhou2016},  which present different types of optimization problem formulations to achieve similar objective functions. 
Profitability and operation possibilities of distributed generation, home batteries and electric vehicles depend on the electricity tariff structure for the point of connection. 
For instance, \cite{Wang2015} presents different electricity markets and electricity tariff structures for HEMS.
\cite{Dietrich2018} provides an economic analysis of storage for self-consumption in Germany and concludes that the cases with high demand and larger PV installations are the only profitable cases.
However, aggregated flexibility services can provide additional value for storage owners and the present paper provides two optimization algorithms that combines both site and their aggregated level solution.

\vspace{-0.35cm}
\subsection{Aggregated flexibility services}
Moreover, aggregated flexibility can be used not only for providing technical services to cope with distribution grid congestion issues \cite{Spiliotis2016} or to increase the grid hosting capacity \cite{Varela2017}, but also can help to improve the efficiency of electricity markets \cite{Eid2016}. 
For instance, 
\cite{Carreiro2017} presents a collection work that represents current trends in energy management systems from the aggregators point of view. 
In addition, \cite{Ottesen2018} proposes a centralised method for aggregators to schedule flexibility in different electricity markets. 
Furthermore, \cite{Burger2017} discusses the potential value that aggregators may provide under different regulatory frameworks and how the inadequacy in regulation can harm other power system objectives.

The most recent works include battery aggregated operation in distribution grids \cite{Kim2017,Resch2019, Alrumayh2019,Hu2019,Zoeller2016,GermanAssociationoftheEnergyandWaterIndustryBDEW2015,Sousa2018,TORBAGHAN201847}. 
\cite{Kim2017} presents an analysis of operating central storage units directly connected to medium-voltage grids to provide power system services. \cite{Resch2019} compares the technical service provision to self-consumption maximization service using a centralised battery at distribution level.
However, \cite{Kim2017,Resch2019} do not consider distributed batteries at prosumer level which can change their operation with an economic incentive mechanism.
Furthermore, \cite{Alrumayh2019} formulates a HEMS capable of providing flexibility to DSOs and \cite{Hu2019} presents a bi-level agent-based optimization algorithm including the DSO operation cost. 
Unfortunately, this algorithm cannot be implemented in some countries. For instance the European Union (EU) electricity market unbundling does not allow to merge DSO and end-users objectives as grid information cannot be shared with aggregators. 
Thus, current studies for the European Union applications assume DSO quantifies the flexibility needed to solve the grid problem in accordance to their operating costs as a separate problem and the aggregator assists the DSO by grid zones as suggested in \cite{Zoeller2016} and \cite{GermanAssociationoftheEnergyandWaterIndustryBDEW2015}.
\cite{Sousa2018} presents a flexibility provision HEMS and it is solved with an heuristic particle swarm optimization algorithm. 
Though, augmented Lagrangian methods facilitate to find global optimal solutions in a reasonable computational time in a distributed manner.
Finally, \cite{TORBAGHAN201847} presents a market based flexibility exchange framework for multiple aggregators competing to solve the same congestion problem. It provides basis to formulate algorithms that are capable to deal with multiple aggregators and the work presented in this paper answers the question regarding re-scheduling FDs where an aggregator needs to compromise by activating a certain flexibility volume. The case of multiple aggregators competing for the same service is out of the scope of this paper.

\vspace{-0.35cm}
\subsection{Alternating direction method of multipliers}
The alternating direction method of multipliers (ADMM) is a decomposition algorithm for distributed convex optimization, but it can be also considered as an heuristic algorithm for solving non-convex problems \cite{boyd2011distributed}. 
Nowadays, ADMM is widely applied in distributed computing environments. 
ADMM can be used to solve problems involving large amount of data and variables in the field of smart grids.
For example, a fully distributed optimal power flow problem is presented in \cite{Erseghe2014}, and \cite{Dvorkin2019} suggests an ADMM approach for a generation investment problem in electricity markets.
In the field of ancillary services, \cite{brooks2015} show a decentralised multi-block ADMM for primary frequency control.

Cloud computing services and big data research can also benefit from ADMM algorithms as they allow to distribute computational power to solve different sub-problems that can be assigned to different processing units \cite{boyd2011distributed}. 
\cite{Liu2015} presents the mathematical formulation of network energy management, robust state estimation and security constrained optimal power flow problems in a distributed manner via ADMM. \cite{Deng2017} formulates the parallel multi-block ADMM for generic problems like exchange problem, $\ell$1-minimization and distributed large-scale $\ell$1-minimization, and tested them on a cloud computing platform. 

For applications regarding energy management in smart grids,
\cite{Sousa2019} reviews the work of different authors on community-based and peer-to-peer (P2P) market mechanisms. 
\cite{Baroche2019} solves a novel cost allocation in P2P electricity markets using the consensus ADMM algorithm.
\cite{Moret2018} formulates a distributed operation of energy collectives using an ADMM algorithm by varying penalty parameter adapted to each case study.
\cite{Liu2019} presents a distributed optimization for community microgrids where each site has its own HEMS. However, the applied ADMM method is very case-sensitive to the augmented Lagrangian penalty parameter for meeting the energy balance constraint. 
Therefore, this paper presents a novel two-steps accelerated method which uses the Proximal Jacobian regularization to find optimization solutions faster without large variations in the objective function.

\vspace{-0.35cm}
\subsection{Summary of contributions}
The novelty of this paper is the formulation and comparison of two flexibility service provision optimization algorithms in centralised and distributed manner. 
First, the centralised algorithm proposes a methodology to determine the battery scheduling for flexibility provision. However, centralised algorithms could have scalability limitations in case of using detailed battery models. 
In such cases, centralised optimization algorithms cannot find reasonably good solutions in less than 10 minutes which is the common time reference to execute energy management decision algorithms in cloud computing platforms.
Therefore, the distributed algorithm presented in this paper consists of a modification of the well known accelerated multi-block Proximal Jacobian ADMM algorithm which includes a novel two-steps dual variable updating in order to avoid high variations in the objective function allowing to update each prosumer site block concurrently.
The case study presented is the evidence for ease of implementation of the distributed algorithm. Additionally, this paper compares centralized and distributed optimization algorithms to highlight the benefits of the novel methodology proposed.

\vspace{-0.3cm}
\section{System Description} \label{sec:SYS}
Aggregators can respond to flexibility requests from BRPs for day-ahead and intraday portfolio optimization or from DSOs to reduce network congestions \cite{VandenBerge2016}.
A FR can be defined as the difference between the baseline and the desired load profile. It can be measured in energy per programming time unit (PTU). 
FR can be for up-regulation which corresponds to an increase in generation or a decrease in consumption. Similarly, the FR for down-regulation can be defined as a decrease in generation or an increase in consumption.
The baseline load profile at transmission level is defined based on market settlements, which cannot be applied at distribution grid level.
Moreover, DSOs, BRPs and aggregators have different portfolios and their baselines cannot be analogous. 
In the present framework, DSO and BRP use aggregator's load baseline as reference for their flexibility settlements. 

This paper assumes the local flexibility market (LFM) framework presented in \cite{Olivella-Rosell2018} where an aggregator is responsible for re-scheduling FDs to meet an external FR without violating the end-user agreements. 
Additionally, aggregator applies a traffic light system as suggested in \cite{GermanAssociationoftheEnergyandWaterIndustryBDEW2015, Olivella-Rosell2018} to solve potential conflicts between simultaneous or even contradictory FR.
Decisions are made centrally using two-way communications, where aggregators send direct control signals and receive metered consumption and status from each FD.
In \cite{Kok2016} it is stated that the architecture used has potential scalability limitations. However, in this paper it is shown that distributed optimization can overcome scalability issues. 

\vspace{-0.3cm}
\section{Site Level Optimization Problem}
The site level optimization problem defined in \eqref{eq:SiteOpt} schedules all FDs by considering battery and PV constraints, forecasted inputs and costs for the site, but not the FR. 
Nomenclature is listed in Appendix \ref{appendix:nomenclature}.
This problem is formulated as an mixed-integer linear programming (MILP) 
as it follows.
Linear objective function~\eqref{eq:SiteOF} represents the cost minimization for prosumer site~\textit{i} including flexibility costs for either using the battery or curtailing the PV if needed.
Additionally, \eqref{eq:electricityBalance} is the electricity balance of site~\textit{i}, \eqref{eq:import_cap}, \eqref{eq:export_cap} limit site import and export capacity and \eqref{eq:eb_binary_var} ensures site~\textit{i} does not import and export electricity simultaneously at period~\textit{t}.
Thus, each prosumer site problem has 72~binary variables including the battery model.
Constraint \eqref{eq:FlexCostDef} defines the flexibility cost function of each site~\textit{i} as the addition of calendar ($\beta_{i,t}^{cal}$) and cycling ageing ($\beta_{i,t}^{cyc}$) costs, and the PV curtailment penalty.
The aggregated result of all prosumer site optimization problems is the load baseline in aggregated flexibility services.

\vspace{-0.2cm}
\begin{subequations}\label{eq:SiteOpt}
	\begin{align}
	\underset{\chi,\zeta}{\text{min}} & \displaystyle\sum_{t \in T}\Big(P_{i,t}^{buy} \chi_{i,t}^{buy} - P_{i,t}^{sell}\chi_{i,t}^{sell} + \zeta_{i,t}^{flex} \Big) & \forall i \label{eq:SiteOF}\\ %Electricity cost 
	\text{s.t.} & \chi_{i,t}^{buy} + \sigma_{i,t}^{dis} + \psi_{i,t} = \chi_{i,t}^{sell} + \sigma_{i,t}^{ch} + W_{i,t}^{l} & \forall i,t \label{eq:electricityBalance} \\
 	&\chi_{i,t}^{buy} \leq \delta_{i,t}^{buy} X_i^{imp} & \forall i,t \label{eq:import_cap} \\
	&\chi_{i,t}^{sell} \leq \delta_{i,t}^{sell} X_i^{exp} & \forall i,t \label{eq:export_cap}\\
	& \delta_{i,t}^{buy} + \delta_{i,t}^{sell} \leq 1 & \forall i,t \label{eq:eb_binary_var} \\
	&	\zeta_{i,t}^{flex} = P_{i,t}^{gen}(W_{i,t}^{gen} - \psi_{i,t}) + \beta _{i,t}^{cyc} +  \beta_{i,t}^{cal} & \forall i,t \label{eq:FlexCostDef}
	\end{align}
\end{subequations}

\vspace{-0.3cm}
\section{Centralised Flexibility Provision} \label{sec:centralisedAlgorithm}
The centralised flexibility service provision algorithm formulated in this section is composed of two consecutive problems: 
\begin{itemize}
	\item  Aggregated level flexibility offer (ALFO) formulation finds the available flexibility without violating local constraints.
	\item Aggregated level flexibility management (ALFM) problem finds the cheapest scheduling of FDs once the aggregator received a FR acceptance from a BRP and/or DSO.
\end{itemize}

\vspace{-0.3cm}
\subsection{Centralised flexibility provision algorithm} 
\begin{figure}
	\centering\includegraphics[width=0.40\textwidth]{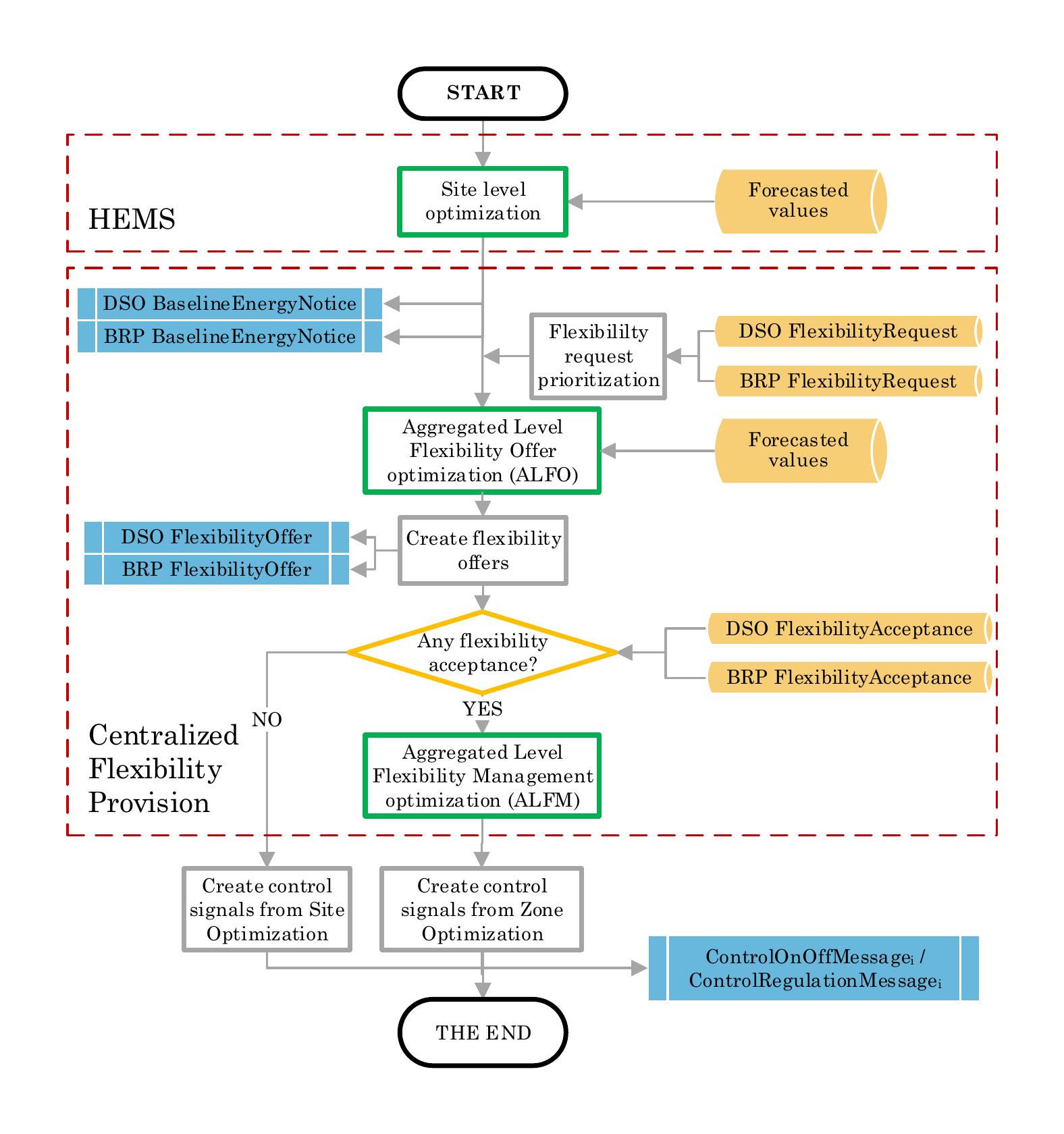}
	\caption{Centralised algorithm flowchart for attending  flexibility requests.}
	\label{fig:flowchart}	
\end{figure}

The algorithm flow diagram for flexibility is shown in Fig.~\ref{fig:flowchart}. The first step optimizes all FDs of each prosumer to reduce energy cost individually following \eqref{eq:SiteOpt}. 
The obtained result is the optimized baseline demand. 
Then, each DSO and BRP receive a notice about the baseline demand of their customers per grid zone and they send a FR based on their projections.
Then, the aggregator executes the aggregated level flexibility offer optimization (ALFO) problem to calculate the available flexibility.
The resulting feasible flexibility is sent back to the BRP or the DSO as a flexibility offer. Each DSO and BRP can decline, or accept the offer fully or in its parts.
In case the offer is completely accepted, the aggregator can use the result of the ALFO optimization to generate the control signals. Otherwise, the aggregated level flexibility management (ALFM) problem re-schedules all FDs at minimum cost to meet the accepted FR. This process schedules all FDs for an optimization planning horizon and it is repeated for every period considering new forecast and FRs.
It is to be noted that the baseline demand could be updated each time new forecasted values are obtained. 
However, it can be convenient to keep a constant baseline within a day to avoid confusions about the reference scenario between aggregator, BRP and DSO. 
\cite{Lloret-Gallego2018} presents example cases of aggregated flexibility services using this architecture.

ALFO and ALFM objective functions and aggregation constraints are \eqref{eq:MaxFlexProblem} and \eqref{eq:MinCostProblem} respectively:

\vspace{-0.3cm}
\subsection{ALFO problem formulation} \label{sec:ALFO}
Objective function~\eqref{eq:OF_MaxFlex} considers the electricity and flexibility costs of each site $i$ over the planning horizon. This problem is formulated as an mixed-integer nonlinear programming (MINLP) as
flexibility costs include quadratic penalty for the flexibility that is not served. 
The penalty is defined as the difference between the total load scheduled in each site optimization problem ($\chi_{i,t}^{tot}$) as per constraint \eqref{eq:ChiTotDef}, and the expected load after applying a FR as per constraint \eqref{eq:WFlexDef}. 
The quadratic penalty ensures that the flexibility provided will follow the shape of the FR.
Otherwise, in case of flexibility scarcity, linear difference would not necessarily generate flexibility provision for all constrained time periods.
Constraints~\eqref{eq:st_MaxFlex_T+} and \eqref{eq:st_MaxFlex_T-} ensure that the activated amount of flexibility is less or equal to the up and down FR respectively.
Additionally, constraints~\eqref{eq:st_MaxPeak+} and \eqref{eq:st_MaxPeak-} are necessary in cases of grid congestions to ensure that the rebound effect is not causing new load peaks before or after the activation of the FR. 

\vspace{-0.3cm}
\begin{subequations} \label{eq:MaxFlexProblem}
	\begin{align}
	\underset{\chi_{i},\zeta_{i}}{\text{min }} & %Electricity cost
	\displaystyle\sum_{i \in I}\displaystyle\sum_{t \in T}\bigg(P_{i,t}^{buy} \chi_{i,t}^{buy} - P_{i,t}^{sell}\chi_{i,t}^{sell} + \zeta_{i}^{flex} + \span \nonumber\\ 
	& P^{penal} \displaystyle\sum_{t\in T^\pm}\Big(W^{flex}_t - \displaystyle\sum_{i \in I}\chi_{i,t}^{tot}\Big)^2\bigg) \label{eq:OF_MaxFlex}\\
	\text{s.t. } & \textstyle\sum_{i \in I}\chi^{tot}_{i,t} \geq 
	W^{flex}_t & \forall t\in T^+ 
	\label{eq:st_MaxFlex_T+}\\
	& \textstyle\sum_{i \in I}\chi^{tot}_{i,t} \leq 
	W^{flex}_t & \forall t\in T^- 
	\label{eq:st_MaxFlex_T-}\\
	& \textstyle\sum_{i \in I}\chi^{buy}_{i,t} \leq \text{max}(W^{flex}_t) & \forall t \label{eq:st_MaxPeak+}\\
	& \textstyle\sum_{i \in I}\chi^{sell}_{i,t} \leq \text{max}(W^{flex}_t) & \forall t \label{eq:st_MaxPeak-}
	\end{align}
\end{subequations}
\vspace{-0.7cm}
\begin{subequations}
	\begin{align}
	\chi^{tot}_{i,t}&=\chi_{i,t}^{buy} - \chi_{i,t}^{sell} \label{eq:ChiTotDef} \\
	W^{flex}_t &= W^{base}_t-FR_t \label{eq:WFlexDef}
	\end{align}
\end{subequations}

\vspace{-0.3cm}
\subsection{ALFM problem formulation}
Once the aggregator estimates the available flexibility, it is communicated as flexibility offers to the BRP or DSO. 
If they accept the offer either partially or entirely, the aggregator can execute the ALFM optimization problem to fulfil it. 
The objective function~\eqref{eq:OF_MinCost} is the same as objective function~\eqref{eq:OF_MaxFlex} but removing the quadratic penalty for not meeting the FR as the ALFO problem ensures there is sufficient flexibility. 
In this MILP optimization problem, constraints~\eqref{eq:st_MinCost_T+} and \eqref{eq:st_MinCost_T-} ensure enough flexibility and the site cost is penalizing the excessive battery usage. 
We can include constraints~\eqref{eq:st_MaxPeak+} and \eqref{eq:st_MaxPeak-} if it is necessary to avoid new undesired load peaks like in Section \ref{sec:ALFO}.

\vspace{-0.4cm}
\begin{subequations} \label{eq:MinCostProblem}
	\begin{align}
	\underset{\chi_{i},\zeta_{i}}{\text{min }} & %Electricity cost
	\displaystyle\sum_{i \in I}\displaystyle\sum_{t \in T}\bigg(P_{i,t}^{buy} \chi_{i,t}^{buy} - P_{i,t}^{sell}\chi_{i,t}^{sell} + \zeta_{i}^{flex}\bigg) \label{eq:OF_MinCost} \span \\
	\text{s.t.} & \textstyle\sum_{i \in I}\chi^{tot}_{i,t} \leq 
	W^{flex}_t & \forall t\in T^+ 
	\label{eq:st_MinCost_T+}\\
	& \textstyle\sum_{i \in I}\chi^{tot}_{i,t} \geq 
	W^{flex}_t & \forall t\in T^- 
	\label{eq:st_MinCost_T-}
	\end{align}
\end{subequations}

The main advantage of this formulation is its simplicity.
However, problems \eqref{eq:MaxFlexProblem} and \eqref{eq:MinCostProblem} may have scalability limitations as mentioned before.
Therefore, this paper explores methods to decompose the problem to improve its computational performance with large-scale implementations.

\vspace{-0.2cm}
\section{Distributed Flexibility Provision} \label{sec:Distrib_FORM}
The distributed flexibility provision algorithm aims to solve the same problem previously presented in Section \ref{sec:centralisedAlgorithm} but using the alternating direction method of multipliers (ADMM) in order to improve computational performance in terms of memory usage and execution time.
The proposed distributed algorithm is based on the optimal exchange problem presented in \cite{boyd2011distributed} but applied in local flexibility markets where each prosumer settles their contribution to the FR in parallel.
Therefore, high performance computers or energy cloud platforms can solve distributed algorithms with less scalability limitations using multi-processor architectures.
Additionally, the distributed formulation can deal with FRs that surpasses the available flexibility. Thus, both ALFO and ALFM are substituted by a single distributed optimization algorithm.

\vspace{-0.3cm}
\subsection{Augmented Lagrangian}
The augmented Lagrangian relaxation ($\mathcal{L}_{\rho}$) is presented in \eqref{eq:lagrangian} and it is equivalent to the previous problems Eqs.~\eqref{eq:MaxFlexProblem} and \eqref{eq:MinCostProblem} relaxing constraints \eqref{eq:st_MaxFlex_T+},  \eqref{eq:st_MaxFlex_T-}, \eqref{eq:st_MinCost_T+}, \eqref{eq:st_MinCost_T-}.
According to \cite{boyd2011distributed}, the regular ADMM algorithm consists of an iterative process to minimize $\mathcal{L}_{\rho}$ and to update the dual variable ($\lambda_{t}^{(k)}$) for each constrained period \textit{t} by giving a fixed penalty parameter ($\rho>0$). In ADMM implementations, it is necessary to identify generic values for penalty parameters ($\rho$) capable of providing satisfying solutions in reasonable computation time for multiple cases. 

\vspace{-0.4cm}
\begin{equation} \label{eq:lagrangian}
\begin{array}{c}
\mathcal{L}_{\rho} = f(x_{i}) + \textstyle\sum_{t \in T^\pm} \lambda_{t}^{(k)}\Big(\chi_{i,t}^{tot} - \dfrac{W^{flex}_t}{N}\Big) + \\
\dfrac{\rho}{2}\textstyle\sum_{t \in T^\pm}
\Big\|\chi_{i,t}^{tot} +  \textstyle\sum_{\substack{j \in I \\ j \neq i }} \chi_{j,t}^{tot,(k)} - (W^{flex}_t))\Big\|^2 
\end{array}
\end{equation}

In the present work, each site \textit{i} can decide individually their contribution to the FR according to its cost function~\eqref{eq:site_cost_ADMM} using ($x_{i}$) as decision variable vector~\eqref{eq:x_def}, the dual variable ($\lambda_{t}^{(k)}$) and the result of other sites~($j\in I, j\neq i$) in the previous step~($k$).

\vspace{-0.4cm}
\begin{subequations}
	\begin{align} 
		f(x_{i}) = & \displaystyle\sum_{t \in T}\Big(P_{i,t}^{buy} \chi_{i,t}^{buy} - P_{i,t}^{sell}\chi_{i,t}^{sell} + \zeta_{i,t}^{flex}\Big) \label{eq:site_cost_ADMM} \\
		x_i = & [\chi_{i}^{tot},\zeta_{i}^{flex}] \label{eq:x_def}
	\end{align}
\end{subequations}

\vspace{-0.3cm}
\subsection{ADMM algorithm modifications}
\cite{Liu2015} and \cite{Ma2016} present two modifications to the original ADMM. 
First, they suggested to update the primal variables concurrently. This allows for a larger problem dimension since it can be solved in a distributed computing system. Also, it reduces the computational cost as the parallelization gains overcome the overhead derived from it.
Second, they introduce the Proximal Jacobian (PJ) regularization term 
$\Big(\dfrac{1}{2} \| (x_{i} - x_{i}^{(k)} ) \|^2_{P_{i}}\Big)$, which preserves parallel updating. 
The norm of this term is defined as $\|x_i\|^2_{P_{i}} = x^T_i P_i x_i$ with $P_i=Id$.
It guarantees strong convexity of the problem and enhances stability. 
The damping parameter $\gamma$ is set to 1.5 as suggested in \cite{Liu2015}.

\vspace{-0.3cm}
\subsection{Penalty parameter and dual variables updating}
Algorithm \ref{alg:ADMM} illustrates the dual update accelerated iteration process divided in two steps: 
\subsubsection{Step 1 - Fast updating}
the early iterations ($k$) are accelerated by varying the penalty parameter $\rho^{(k)}$ according to \eqref{eq:rho_update_1} as per the approach adapted by \cite{boyd2011distributed}. $\tau^{incr}$ and $\tau^{decr}$ are increasing and decreasing factors in order to speed up the algorithm.
$\rho^{(k)}$ is used to update dual variables $\lambda_t^{(k+1)}$ according to \eqref{eq:dual_update_1}, and
\eqref{eq:primal_error} and \eqref{eq:dual_error} are the primal and dual error definitions respectively. 
$r^{(k)}_t$ is positive when the set of prosumer sites \textit{I} cannot provide enough up-regulation to meet the FR at iteration $k$ and dual variable $\lambda_{t}^{(k)}$. Otherwise,  $r^{(k)}_t$ is negative.

\vspace{-0.2cm}
\begin{equation} \label{eq:rho_update_1}
\rho^{(k+1)} := 
\begin{cases}
\tau^{incr}\rho^{(k)} & \text{if } \|r^{(k)}\|_2 > \mu \|s^{(k)}\|_2 \\
\rho^{(k)}/\tau^{decr} & \text{if } \|s^{(k)}\|_2 > \mu \|r^{(k)}\|_2 \\
\rho^{(k)} & \mbox{otherwise}, \\
\end{cases}
\end{equation}

\vspace{-0.2cm}
\begin{subequations}
	\begin{align}
	\lambda_{t}^{(k+1)}&:=\lambda_{t}^{(k)} + \gamma\rho^{(k)} r^{(k)}_t & \forall t \in T^\pm \label{eq:dual_update_1} \\
	r^{(k)}_t & = \textstyle\sum_{i\in I}\Big(\chi_{p,t}^{tot,(k)}-W_t^{flex}\Big) & \forall t \in T^\pm \label{eq:primal_error} \\
	s^{(k)}_t & = r_t^{(k)} - r_t^{(k-1)} & \forall t \in T^\pm \label{eq:dual_error}
	\end{align}
\end{subequations}

\subsubsection{Step 2- Soft updating} 
the dual update changes to \eqref{eq:dual_update_2} at iteration $k^*$ once the primal error is equal or less than threshold value like 5$\%$ of the FR. 
Then, it starts collecting accumulated primal error over iterations from $k^*$ to $k$ and includes integration and dual residual regulation parameters $K^i$ and $K^d$ inspired in classic control theory \cite{kuo1995automatic}. 
They allow to better regulate the dual variables update and damp oscillations in the error and objective function along the iterative solution process.
However, at this moment it is unclear how to tune the parameters to accelerate the convergence.
It is to be noted that penalty parameter $\rho^{(0)}$ is the initial value. 

\vspace{-0.3cm}
\begin{align} \label{eq:dual_update_2}
\lambda_{t}^{(k+1)} 
& := \lambda_{t}^{(k)} + \gamma\rho^{(0)} r^{(k)}_t + & \nonumber \\
& K^i\displaystyle\sum_{i=k^*}^{k}r^{(i)}_t + K^d s^{(k)}_t & \forall t \in T^\pm
\end{align}

The algorithm stops when the norm of the primal and dual residuals are both smaller than some given thresholds ($\epsilon^{pri}$ and $\epsilon^{dual}$ respectively). Additionally, a maximum number of iterations $k^{max}$ and computational time ($CT^{max}$) are specified as well. Depending on the requirements of the case, it might be interesting to compute as many iterations as possible in a pre-specified time, or on the contrary run for as long as needed the algorithm until we reach a precision requirement. Both scenarios are tested in Section \ref{sec:RES} experiments.

\begin{algorithm} \small
	\SetAlgoLined
	Initialize: $x_{i}^{(0)}, \lambda_{t}^{(0)}, \rho^{(0)}>0 $ \;
	\KwIn{$K^i, K^d, \epsilon^{pri}, \epsilon^{dual}, \tau^{incr}, \tau^{decr}, CT^{max}, k^{max}, W^{flex}$}
	
	\While{$\epsilon^{pri}>\|r_t^k\|_2$ and $\epsilon^{dual}>\|s_t^k\|_2$}{
		\For{i=1,2,...,I}
		{
			($x_i$ is updated \textbf{concurrently})\;
			$x_{i}^{(k+1)}$ := $\underset{x_i}{\text{argmin}}$ $\mathcal{L}_{\rho}(x_i, {x_j^k}_{j\neq i},\lambda_{t}^{k},\rho^{(k+1)}) + \dfrac{1}{2} \| (x_{i} - x_{i}^{(k)} ) \|^2_{P_{i}}$\;
			s.t. \text{Site $i$ constraints:} (\ref{eq:electricityBalance})\eqref{eq:import_cap}\eqref{eq:export_cap}\eqref{eq:eb_binary_var}
		}
		{
			Update dual and penalty variables \;	
		}
		\eIf{$\|r_t^{(k)}\|_2 > 0.05\|FR_t\|_2$}{
			Fast update: $\rho^{(k+1)}$ according to \eqref{eq:rho_update_1}\;
			Update $\lambda_{t}^{(k+1)}:=\lambda_{t}^{k}+\gamma \rho^{(k+1)} r_t^{(k+1)},$ $\forall t \in T^{\pm}$\;
		}{
			Soft update: $\lambda_{t}^{(k+1)}$ according to \eqref{eq:dual_update_2}
			%Update $\lambda_{t}^{(k+1)} := \lambda_{t}^{(k)} + \gamma\rho^{(0)} r^k_t + K^i\displaystyle\sum_{i=k^*}^{k}r^i_t + K^d s^k_t$\;
		}{
			Update $\|r_t^{(k)}\|_2$, $\|s_t^{(k)}\|_2$, $k=k+1$
		}
	}
	\caption{Two-steps Fast-PJ-ADMM for optimal flexibility provision.}
	\label{alg:ADMM}
\end{algorithm}

\vspace{-0.3cm}
\section{Case study} \label{sec:case_study}
The case study considered for analysis consists of 100~domestic houses which have photovoltaic panels installed and their historic measurement data are available through the Dataport \cite{PecanStreetInc.}. 
The household consumption data and PV generation data considered for the case study are for the date June the 28th of 2018 because the consumption and PV generation were significantly high.
The electricity price is from the Spanish day-ahead market price for the same day in combination with the Spanish two-periods grid tariff which has valley hours from 11~pm until 13~pm on the next day (14~hours) and the rest of the hours are considered as peak hours for summer period.
Metering and forecasted values have hourly resolution and the optimization horizon is one day.
The optimization process is assumed to be executed at 11:45~pm each day and takes decisions based on the load and PV power production forecast for each site.
For simplicity, the study considers only one FR which is for one period (1~hour).

The simulation set-up will be complete by adding a battery to each household, which is the flexibility source and the battery model is explained in the Appendix \ref{appendix:battery_model}.
Battery parameters for the case study are the investment cost which is EUR~3,000 for the first site and it is increased in steps of 1$\%$ for the successive sites.
Moreover, batteries' maximum charging and discharging power are 3.8~kW and capacity is 10~kWh for all batteries.
Battery converter parameters and efficiency are taken from technical data sheet of SMA-SBS3.7-10 converter by assuming the average operational voltage is $V_{DC}=550$~V \cite{SMASolarTechnologyAG}. 
The diversity in the battery technology and their ageing in the considered population of 100 batteries are also represented by their efficiency.
The battery efficiency for the first site is considered as  98$\%$ and the battery efficiency in the successive households are reduced in steps of 0.1$\%$.
The values used for constant voltage charging parameter and calendar ageing related parameters are as follows $W^{bat}_i=0.2$, $S^{LT}_i=10$~years, $S^0_i=0.3$ and $S^{SOC}_i=1.7$ according to \cite{Hentunen2018,Ecker2014a}, and they are the same for all battery units. 
The cycle ageing model considers the battery as 10~segments.

The FR in this case study is formulated to reduce the evening peak at 8~pm according to the local DSO needs. 
Though the household consumption shows a peak at 10~pm, the network congestion can happen at hours other than hours at which the households portfolio shows a peak. 
Therefore, the present case study does not cut the portfolio peak as it is not the purpose of the present flexibility service provision.

\vspace{-0.2cm}
\section{Results} \label{sec:RES}
\subsection{Site level optimization}
In site level optimization, every individual prosumer is optimized to achieve low energy cost independent from others. The aggregation of each site’s net load after optimization of the case study forms the baseline load and it is shown in Fig.~\ref{fig:max_flex_results} 
In site optimization, results show that batteries are partially charged during the afternoon to store the PV power production surplus and they are discharged during the evening. 
This phenomenon is due to calendar ageing penalty, batteries tend to charge at the latest possible low priced period to reduce the time between charge and discharge. In addition, the cycling and calendar ageing factors prevent to charge and discharge batteries for arbitrage as the economic margin does not surpass the battery degradation costs.
It is noticeable that some PV systems are oversized. As there is not enough load during sunny hours or battery capacity, the baseline load shows some periods with net aggregated export of energy.

\vspace{-0.3cm}
\subsection{Centralised flexibility provision}
\begin{figure}
	\centering
	\includegraphics[width=0.48\textwidth]{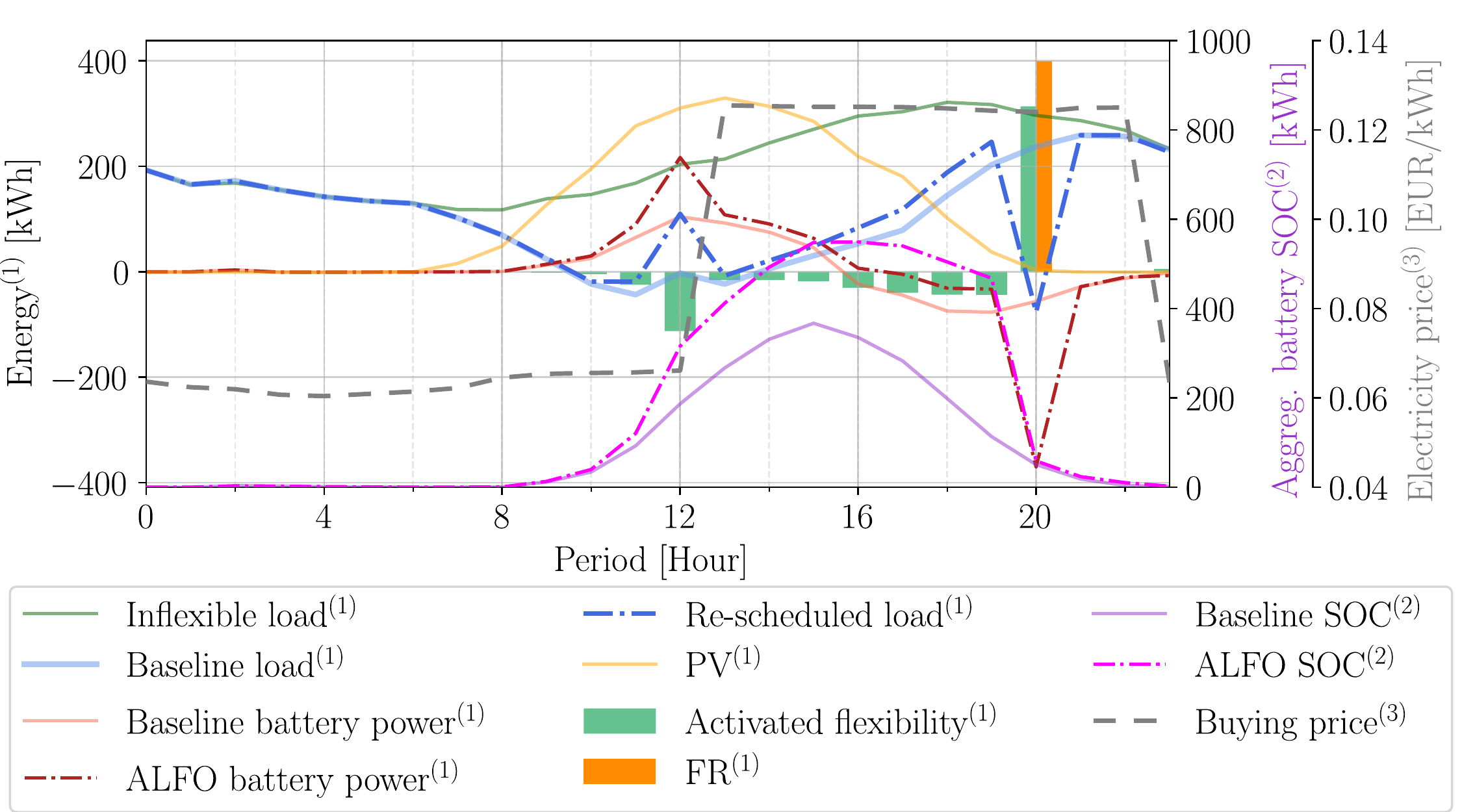}
	\caption{ALFO problem results of re-scheduling of 100 sites and batteries under a FR of 400~kWh.}
	\label{fig:max_flex_results}
\end{figure}

This section shows the results of centralised ALFO and ALFM algorithms for the described case study. 
The available flexibility is calculated by the ALFO problem. 
The FR is for 400~kWh at 8~pm and the households portfolio provides 308.86~kWh as the maximum available flexibility. 
Aggregator cannot provide more flexibility as some battery discharges are already scheduled by the site optimization problem.
Fig.~\ref{fig:max_flex_results} shows the increase in battery charging during the hours before FR to charge all batteries sufficiently for latter discharge at 8~pm during FR. 

\begin{figure}
	\includegraphics[width=0.48\textwidth]{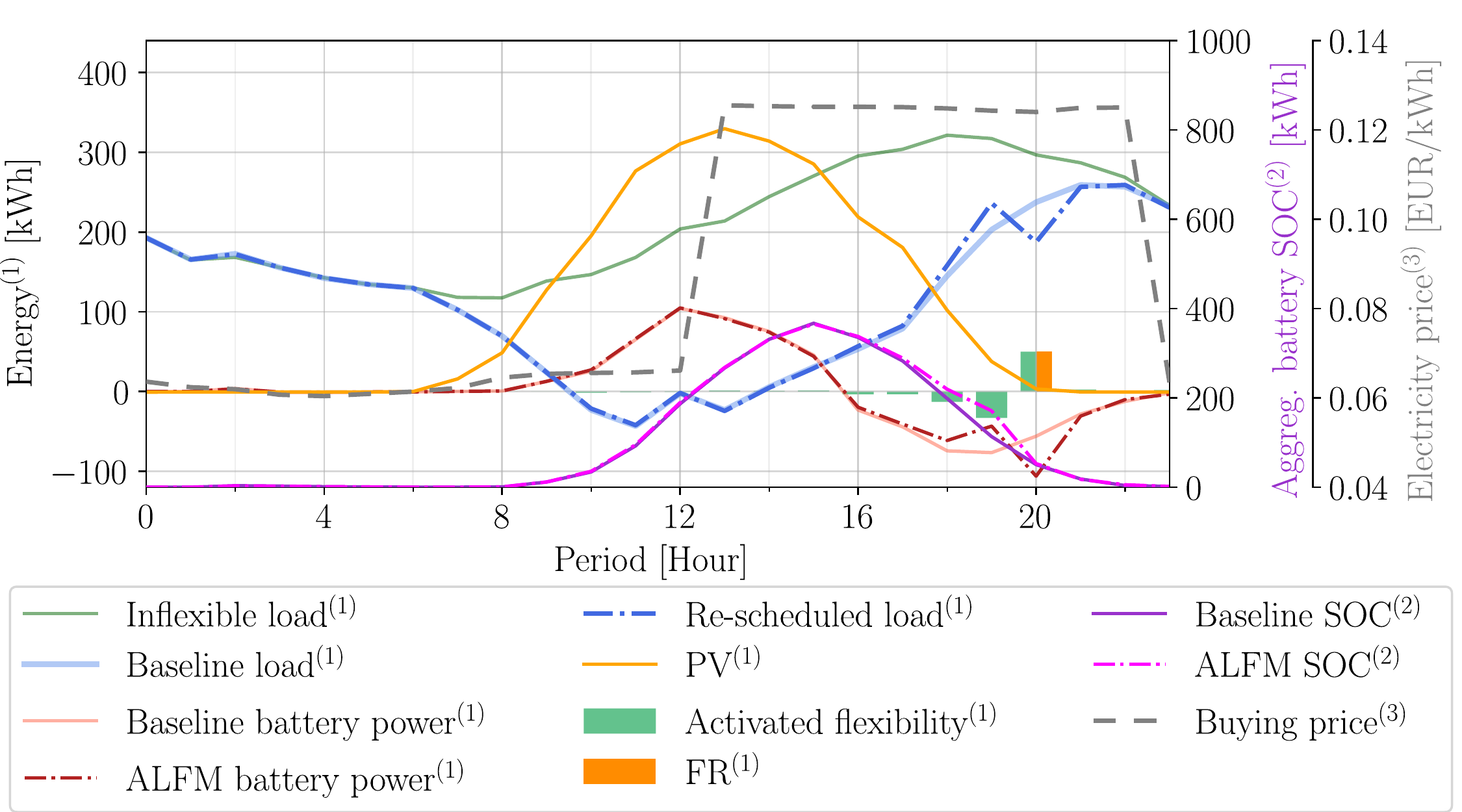}
	\caption{ALFM problem results under a FR of 50~kWh of 100~sites.}
	\label{fig:min_cost_results}
\end{figure}

With the same scenario, if FR is only 50~kWh (lower than the available maximum flexibility), the ALFM problem starts to re-schedule batteries to meet the new FR. 
The complexity of the case study is the significant number of possibilities to attend the request as the FR is close 17$\%$ of the portfolio available flexibility.
Fig.~\ref{fig:min_cost_results} shows the new feasible solution for the FR at minimum cost for all sites.

\vspace{-0.2cm}

\vspace{-0.2cm}

\subsection{Distributed flexibility provision}
\begin{figure}
	\centering
	\includegraphics[width=0.48\textwidth]{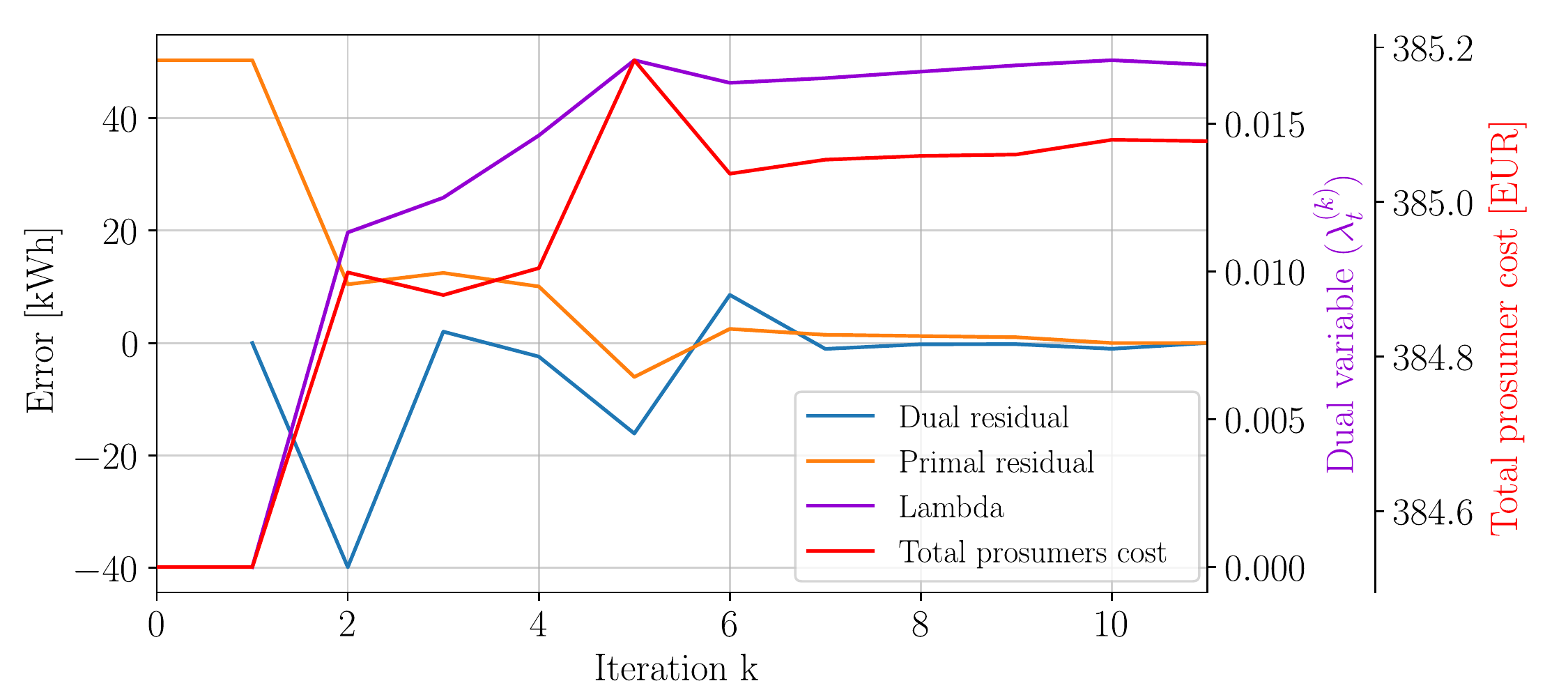}
	\caption{Comparison of errors, total prosumers costs and dual variable per iteration of Algorithm \ref{alg:ADMM} with $K^i=2\cdot 10^{-4}$ and $K^d=5\cdot 10^{-7}$.}
	\label{fig:lambda_ADMM}
\end{figure}

The proposed distributed ADMM Algorithm~\ref{alg:ADMM} is tested with the same case study compared with centralised algorithm.
Fig.~\ref{fig:lambda_ADMM} shows the primal and dual errors, dual variable $\lambda_t^{(k)}$ and the total prosumer cost converge for 11~iterations. 
The rate of change of primal error in successive iterations is high till the iteration $k=5$. 
After that, the error variation is low. This phenomenon is due the fast dual variable updating according to \eqref{eq:dual_update_1} when error is lower than 5$\%$ of FR. 
Thereafter, the soft updating is activated and it changes error rate smoothly by avoiding large variations.
The initial value of $\lambda_{t}^{(0)}$ at zero provides a solution which corresponds to the site optimization result. When Algorithm~\ref{alg:ADMM} updates $\lambda_t^{(k)}$, the portfolio tends to increase flexibility provision and the primal error decreases.
The solution is found in less than 5~minutes and the memory usage is very low (around 200 MB) as each iteration is a separated optimization problem per site and the memory contents are flushed after each iteration.

\vspace{-0.2cm}
\subsection{Scalability analysis}
\begin{figure}
	\centering
	\includegraphics[width=0.48\textwidth]{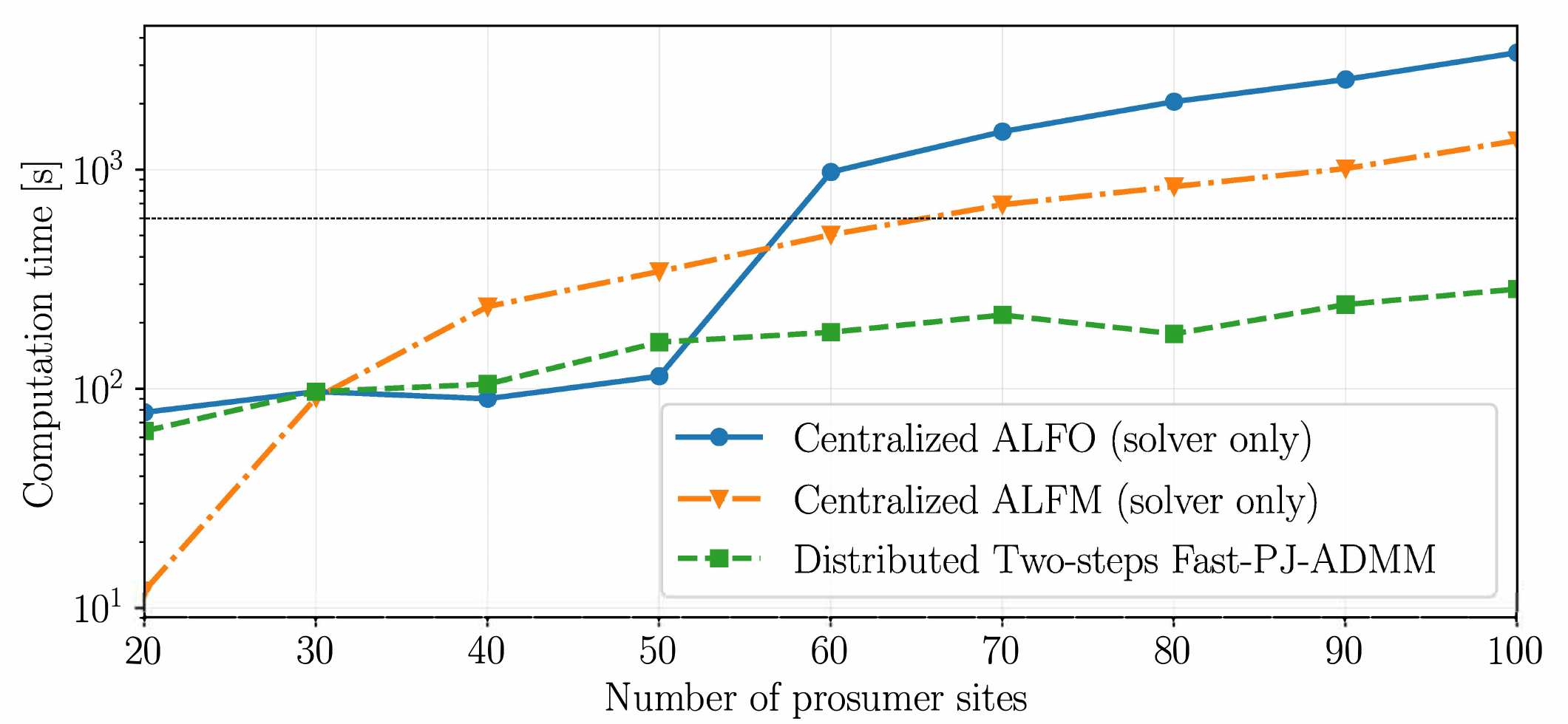}
	\caption{Computational cost comparison of ALFO, ALFM and Fast-PJ-ADMM Algorithm \ref{alg:ADMM}. Horizontal dashed line highlights 10 minutes computation time threshold.}
	\label{fig:simulation_cost_log}
\end{figure}

Regarding scalability limitations, the centralised algorithms using the Gurobi solver with the branch-and-bound and the dual simplex algorithms, they consume the maximum available RAM memory of 16 GB for the 100~sites case study and takes approximately 1~hour to solve the ALFO problem and around 20 minutes for the ALFM problem on high performance computer with AMD Ryzen Threadripper 16 Core Processor running at 3.5~GHz. 
The ALFO problem takes more time due to the quadratic flexibility penalty term in the objective function. 
Fig.~\ref{fig:simulation_cost_log} shows that the Fast-PJ-ADMM Algorithm~\ref{alg:ADMM} which finds solutions at similar duration in comparison with ALFM and ALFO problems in the case with 50~sites but centralised algorithms take between 5 and 12 folds more time for the 100~sites case.
Therefore, there is a scalability limit to solve large-scale flexibility problems with complicating constraints~\eqref{eq:st_MinCost_T+} and \eqref{eq:st_MinCost_T-}, and detailed battery models using centralised algorithms. 

\vspace{-0.2cm}
\subsection{Distributed algorithm acceleration comparison}
This section discusses around four versions of the ADMM algorithms previously presented in the literature namely Regular (orignal algorithm \cite{boyd2011distributed}), Fast (acceleration via penalty parameter \cite{boyd2011distributed}), PJ (parallelized, regularized version \cite{Ma2016}), and Fast-PJ (combination of acceleration, parallelization, and regularization), and the novel version Two-steps Fast-PJ Algorithm \ref{alg:ADMM} by comparing the way penalty parameter ($\lambda_t^{(k)}$) is updated for the optimal flexibility exchange problem in each algorithm. 
Therefore, it is possible to see the impact of acceleration, parallelization, regularization and the two-step algorithm modifications.
All accelerated algorithms begin with the same penalty parameter value ($\rho^{(0)}=10^{-4}$) and they continue with the soft update when $\|r_t^{(k)}\|_2 \text{ and } \|s_t^{(k)}\|_2\leq0.05$~kWh. 

\begin{figure}
	\includegraphics[width=0.48\textwidth]{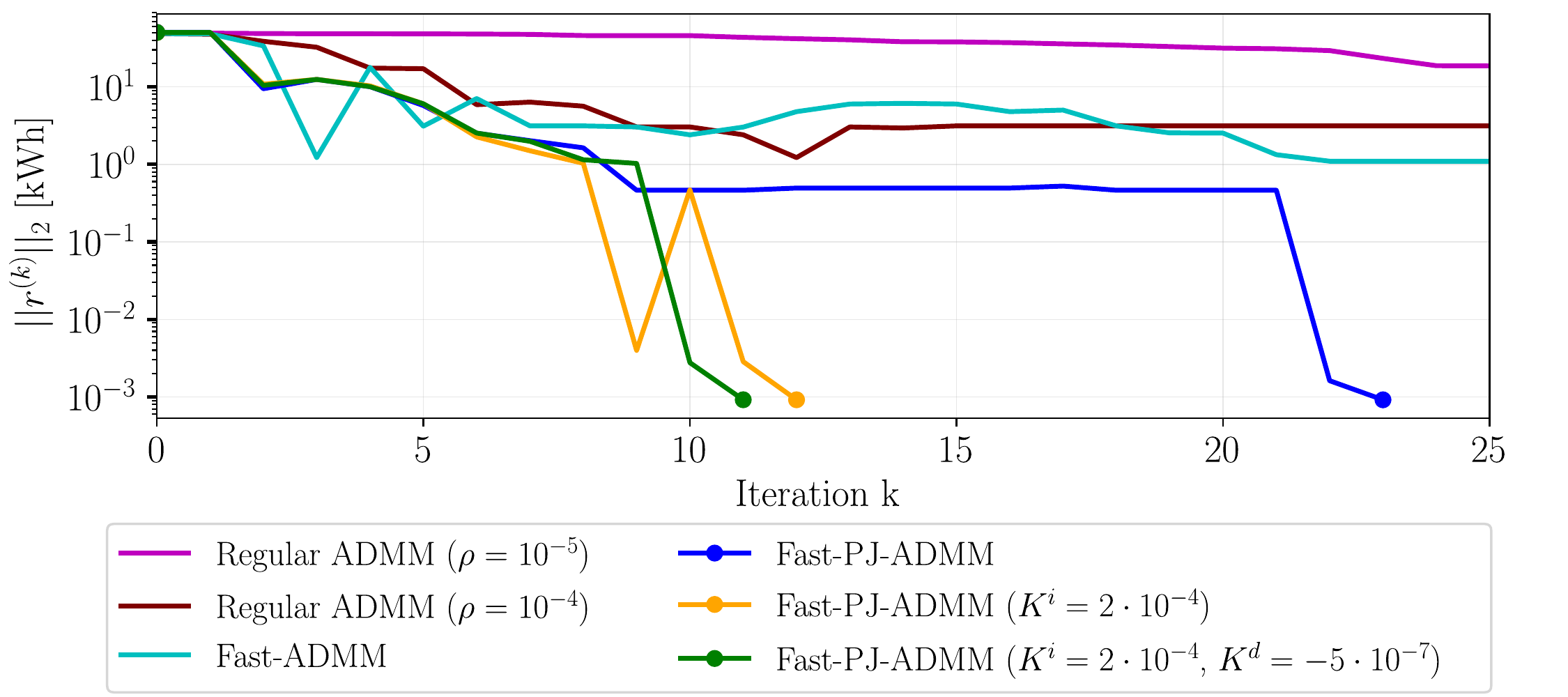}
	\caption{Primal error comparison of different acceleration algorithms under a FR at 8~pm of 50~kWh in a portfolio of 100~sites including the soft updating in all algorithms when $r^{(k)}=2.5$~kWh. Markers indicate when algorithms meet $\epsilon^{pri}=\epsilon^{dual}=10^{-3}\%$.}
	\label{fig:error_ADMM}
\end{figure}

Fig.~\ref{fig:error_ADMM} shows a comparison of absolute primal error values change per iteration of different distributed optimization algorithms. 
It is to be noted that the absolute error value hide cases when primal errors varies between positive and negative values as shown in Fig.~\ref{fig:lambda_ADMM}.
Although they are reaching the same optimal cost,
the regular ADMM with $\rho=10^{-5}$ takes 20~iterations to find a solution with primal error $r_t^{(k)}=20$~kWh, and 200~iterations for 1$\%$~error. In case of increasing the penalty parameter to $\rho=10^{-4}$, the regular ADMM performance differs from the previous case near sub-optimal solutions from \textit{k=4}. This phenomenon is due to the drastic change in dual update. 
The regular ADMM and all the following algorithms proved to provide better solutions if they use an initial value for dual variable as $\lambda_t^{(0)}=0$. Thus, the performance of the algorithm differs from no flexibility provision to the optimal battery schedule to provide full FR.

The Fast-ADMM 
changes the penalty parameter according to \eqref{eq:rho_update_1} using the following acceleration parameters: $\tau^{incr}=1.5$, $\tau^{decr}=2$ and $\mu=2$. 
In all accelerated algorithms, dual variable $\lambda_t$ changes at a higher rate between successive iterations until iteration \textit{k=5} where they reach a primal error equal or lower than 5$\%$ of FR ($\|r_t^{(k)}\|_2=2.5$~kWh). 
Thereafter, their performance differs from each other around the optimal solution. 
The inclusion of the PJ regularization term in the Algorithm \ref{alg:ADMM} according to \cite{Liu2015} allows to stabilize the objective function and it finds an optimal solution with 2$\%$ error in 9~iterations. 
Additionally, it finds an optimal solution ($\epsilon^{pri}<10^{-3}\%$) in 24~iterations if parameters are $\rho^{(0)}=10^{-4}$ and $K^d=K^i=0$ during the soft update. 
In contrast to the previous algorithms, Fast-PJ-ADMM Algorithm~\ref{alg:ADMM} with $K^i=2\cdot 10^{-4}$ and $K^d=-5\cdot 10^{-7}$ finds the optimal solution in 11~iterations. 
The effect of $K^i$ is to faster approach to the objective function expected value and $K^d$ smooths variations.

\begin{figure}
	\centering
	\includegraphics[width=0.48\textwidth]{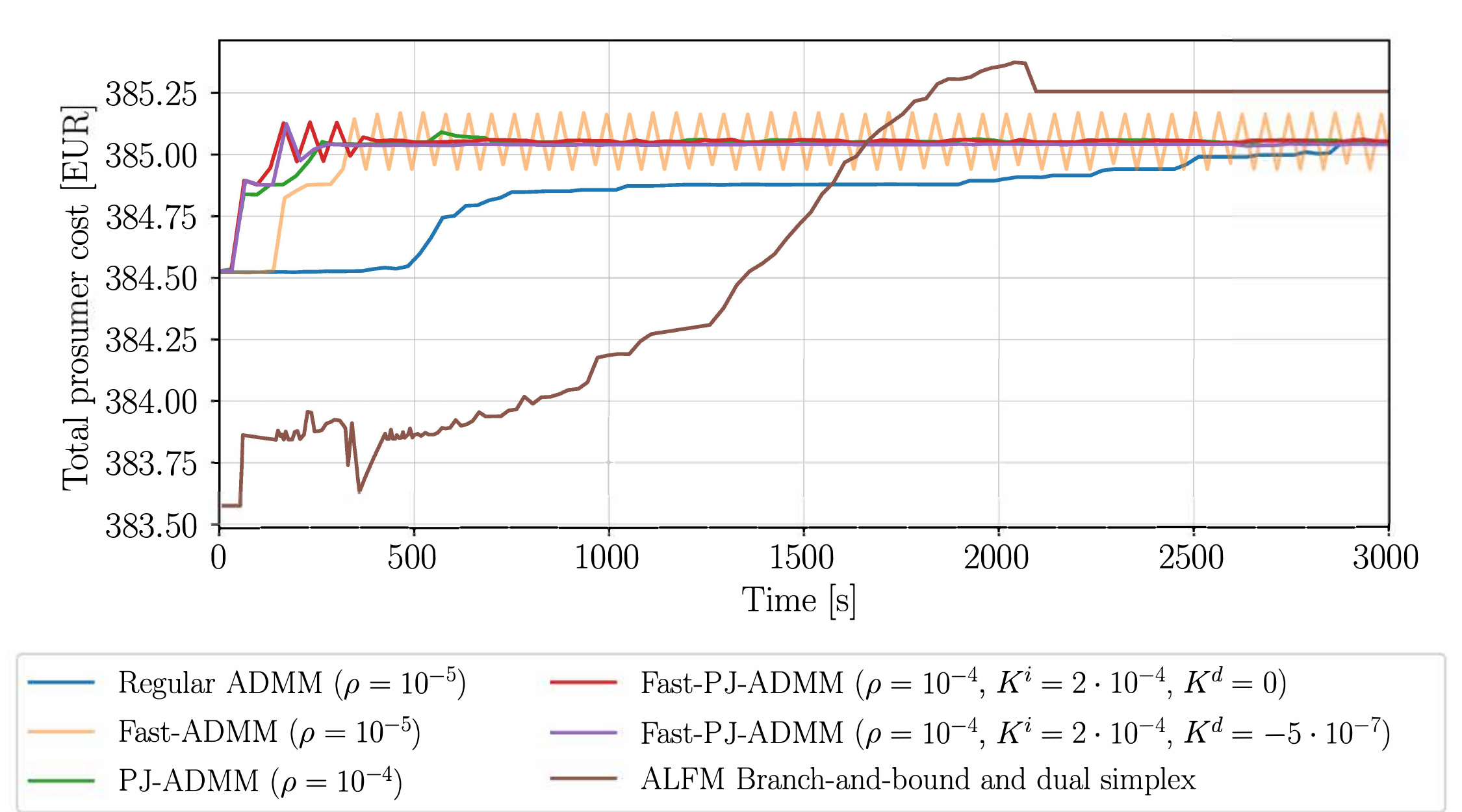}
	\caption{Prosumer total cost over time of centralised and distributed algorithms with FR=50~kWh and $\epsilon^{pri}=\epsilon^{dual}=10^{-5}\%$.}
	\label{fig:objfunc_FR=50kWh}
\end{figure}

Fig.~\ref{fig:objfunc_FR=50kWh} shows the total prosumer cost variation over the resolution time of centralised and multiple distributed algorithms. 
All methods find very similar solutions but they approach to the objective value differently. 
The centralised ALFM algorithm takes 2,000 seconds before reaching a feasible solution. In contrast, most of distributed algorithms find a suitable solution in less than 200 seconds. Notice the Regular ADMM takes 3,000 seconds to reach the optimal value and the Fast ADMM varies around the optimal solution. The PJ regularization term reduces variations and the $K^i$ and $K^d$ parameters allow to accelerate the path to the optimal objective cost without creating variations in the objective function. 

\vspace{-0.2cm}

\section{Conclusions} \label{conclusions}
The present paper presents a novel formulation to optimize the operation of distributed storage units behind-the-meter to provide flexibility services  to a balance responsible party or distribution system operator. In this context, an aggregator manages a group of prosumers with storage units who are willing to participate in the local flexibility market.
In addition, this paper includes the decomposed optimization problem formulation for large-scale portfolios using a modified accelerated PJ-ADMM algorithm divided in two steps: fast and soft dual variable updating. 
The fast and soft updating accelerate the iterative process reducing variations in the dual variable and objective functions. 
The soft update might be relevant for case studies with binary variables as dual errors can be zero although the dual variable changes.
Case study results show that this formulation is best suited for large scale implementations as it can find aggregated optimal solutions faster by considering local constraints and prices with the appropriate parameters tuning.
The results also highlight that the centralised and distributed methods find very similar solutions but the distributed one can overcome the scalability limitations. For instance, the case study shows a break-even point at 50 prosumer sites when the distributed algorithm is less computationally demanding than the centralised.
In future, the performance of the proposed decomposition algorithm can be tested by including other flexibility device models such as electric water heaters or electric vehicles which may increase the complexity of their aggregated optimization.
Additionally, the two-steps Fast-PJ-ADMM algorithm can be applied for a peer-2-peer local market to find optimal exchanges. 

\vspace{-0.2cm}
\section*{Acknowledgement} 
This work has been supported by the INVADE H2020 project (2017--2019), which has received funding from the European Union's Horizon 2020 research and innovation program under Grant Agreement No. 731148 and by the InnoEnergy PhD School. The authors thank Pecan Street, Inc. for allowing access to his DataPort database. 
E. Prieto is lecturer of the Serra H\'{u}nter programme.

\vspace{-0.2cm}

\ifCLASSOPTIONcaptionsoff
  \newpage
\fi

\bibliographystyle{IEEEtran}
\bibliography{references}

\newpage

\appendices
\section{}
\label{appendix:nomenclature}
\printnomenclature[2cm]

\section{Stationary battery model} \label{appendix:battery_model}
This appendix provides the stationary battery model used in the present paper. The elemental battery state-of-charge (SOC) evolution constraint of prosumer site \textit{i} is shown in \eqref{eq:socev}. 
Battery SOC has upper ($O_i^{max}$) and lower boundaries ($ O_i^{min}$), and charging and discharging power limits are ($Q_i^{ch}$) and ($Q_i^{dis}$).

\vspace{-0.3cm}
\begin{subequations}\label{eq:elemental_battery}
	\begin{align}
	\sigma _{i,t}^{SOC} =  &  \sigma _{i,t-1}^{soc} + \sigma _{i,t}^{ch} A_i^{bat,ch} a^{inv,ch} \span \nonumber \\ 
	& - \frac{\sigma _{i,t}^{dis}}{A_i^{dis} a^{inv,dis}}  & & \forall i,t \label{eq:socev}\\
	& O_i^{min} \leq \sigma _{i,t}^{SOC} \leq O_i^{max} & &  \forall i,t \label{eq:socminmax}\\
	& \sigma _{i,t}^{ch} \leq Q_i^{ch} \delta_{i,t}^{bat}& & \forall i,t \label{eq:chmax}\\
	& \sigma _{i,t}^{dis} \leq Q_i^{dis}(1-\delta_{i,t}^{bat}) & & \forall i,t \label{eq:dismax}
	\end{align}
\end{subequations}
Because batteries are expensive and suffer from higher rate of degradation under heavy stress \cite{Ecker2014a}, an advanced battery model has been developed in order to make more accurate decisions. 
The developed battery model is formulated by including battery degradation cost in such a way to reflect real costs of operation as closely as possible to reality while maintaining the computational burden at an acceptable level.
The battery model has 4 main attributes in addition to the simple model which are cycle and calendar degradations, power limitations when approaching fully charged and fully discharged states to avoid reaching the voltage limits, and piecewise linearized inverter efficiency. The following section describes these attributes in detail. 

\subsubsection{Cycle degradation}
The most common stationary batteries at end-user level today are lithium ion batteries, typically li-ion nickel-manganese-cobalt (LI-NMC) batteries. 
The degradation factors of such batteries are predominantly depending on charge-discharge cycle depth during operation.
Therefore, the lifetime of these batteries depends on the depth and number of cycles.
In addition, shallow cycles have significantly lower degradation cost than deep ones. A detailed cycle degradation model is presented
in \cite{Xu2018}. The cycle degradation model factorizes the cycle depth and accounts degradation as a cost of discharging the battery with a certain depth. 
In order to keep track of the depth-of-discharge, the battery model adds virtual segments indexed by $j$ as presented in \cite{Xu2018}. 
\eqref{eq:segsocev} tracks the segmented SOC evolution given segmented variables, whereas \eqref{eq:segmaxsoc} restrains the maximum energy per segment.
\eqref{eq:segch} and \eqref{eq:segdis} sums the power in all segments $j$ to equal the original variables. 
Finally, \eqref{eq:cycageing} calculates the degradation cost as function of discharge power.

\subsubsection{Calendar ageing}
The calendar ageing is modelled as a function of SOC dependent cost per time period, as shown in \eqref{calageing}. 
The core idea is that calendar based degradation cost increases with higher SOC and it incentives the battery to stay at a low SOC when not utilized. The tuning factors $S^0_i$ and $S^{SOC}_i$ implicate how much the calendar ageing depends on SOC as described in \cite{Hentunen2018}.

\subsubsection{Constant-voltage charging/Constant-current discharging}
The constant-voltage charging and constant-current discharging regions of a battery does not apply to the full SOC area of a battery. \eqref{eq:CVch} and \eqref{eq:CVdis} reduces the allowed charging and discharging power when approaching the maximum and minimum energy levels respectively.

\vspace{-0.3cm}
\begin{subequations}
	\begin{align}
	\sigma _{i,t,j}^{seg,SOC} = \sigma _{i,t,j}^{seg,ch} A^{ch} a^{inv,ch}(\sigma _{i,t,j}^{seg,ch}) - \span \nonumber \\
	\frac{\sigma _{i,t,j}^{seg,dis}}{A^{dis}a^{inv,dis}(\sigma _{i,t,j}^{seg,dis})} + \sigma _{i,t - 1,j}^{seg,SOC}  & & \forall i,t \label{eq:segsocev}\\
	\sigma_{i,t, j}^{seg,SOC} \leq O_{i,j}^{seg,max}  & & \forall i,j,t \label{eq:segmaxsoc}\\
	\sigma _{i,t}^{ch} = \textstyle\sum_{j \in J}\sigma _{i,t,j}^{seg,ch}  & & \forall i,t \label{eq:segch}\\
	\sigma _{i,t}^{dis} = \textstyle\sum_{j \in J}\sigma _{i,t, j}^{seg,dis}  & & \forall i,t \label{eq:segdis}\\
	\beta _{i,t}^{cyc} = \textstyle\sum_{j \in J} C_{i,j} \sigma _{i,t,j}^{seg,dis}  & & \forall i,t \label{eq:cycageing} 
	\end{align}
\end{subequations}
\begin{equation}
\beta _{i,t}^{cal} = \frac{C_i^{bat}}{S_i^{LT}}\cdot \Big(S_i^{0}+\frac{1}{2}S_i^{SOC}\cdot (\sigma _{i,t}^{SOC}+\sigma _{i,t-1}^{SOC})\Big) \forall i,t \label{calageing}
\end{equation}

\begin{subequations}
	\begin{align}
	\sigma _{i,t}^{ch} \leq \frac{O_i^{max}-\sigma _{i,t-1}^{SOC}}{(1+W^{bat}_{i})}&  & \forall i,t \label{eq:CVch} \\
	\sigma _{i,t}^{dis} \leq \frac{\sigma _{i,t-1}^{SOC}-O_i^{min}}{(1+W^{bat}_{i})} &  & \forall i,t \label{eq:CVdis}
	\end{align}
\end{subequations}

\subsubsection{Piecewise linearized inverter efficiency}
The total storage system efficiency is a combination of two factors, the inverter efficiency ($a^{inv,ch}$) and the battery efficiency ($A^{bat,ch}$). A piecewise linearized approach is chosen in order to capture the power dependency of inverter efficiency. 
At low input power inverter efficiency is very low, on the other hand the efficiency reaches 98\% at high input power. 
The piecewise linearization is modelled using a special order sets of type 2 (SOS2) approach to approximate the non-linear dependency on input power. This approach adds four additional binary variables per time step (two for charging, two for discharging) to the problem, and is one of the preferred methods first developed in \cite{Beale1976}.

\end{document}